\documentclass[lettersize,journal]{IEEEtran}
\usepackage{amsmath,amsfonts}
\usepackage{algorithmic}
\usepackage{algorithm}
\usepackage{array}
\usepackage[caption=false,font=normalsize,labelfont=sf,textfont=sf]{subfig}
\usepackage{textcomp}
\usepackage{stfloats}
\usepackage{url}
\usepackage{verbatim}
\usepackage{graphicx}
\usepackage{cite}
\usepackage{amsthm,amssymb}
\usepackage{mathrsfs}
\usepackage{xcolor}

\begin{document}

\title{ Joint Low-Rank and Sparse Bayesian Channel Estimation\! for\! Ultra-Massive\! MIMO\! Communications}

\author{Jianghan~Ji, Cheng-Xiang~Wang,~\IEEEmembership{Fellow,~IEEE}, Shuaifei~Chen,~\IEEEmembership{Member,~IEEE}, 
\\
Chen~Huang,~\IEEEmembership{Member,~IEEE}, Xiping~Wu,~\IEEEmembership{Senior Member,~IEEE},
and Emil Bj{\"o}rnson,~\IEEEmembership{Fellow,~IEEE}\vspace{-2em}

\thanks{This work was supported by the National Natural Science Foundation of China (NSFC) under Grants 62401643, the China Postdoctoral Science Foundation under Grant 2024M752439, and the Research Fund of National Mobile Communications Research Laboratory, Southeast University, under Grant 2025A05. E.~Bj{\"o}rnson was supported by the Grant 2022-04222 from the Swedish Research Council. (Corresponding authors: Cheng-Xiang Wang and Shuaifei Chen.)

J. Ji is with the National Mobile Communications Research Laboratory, School of Information Science and Engineering, Southeast University, Nanjing 211189, China (e-mail: ji\_jh@seu.edu.cn).

C.-X. Wang and X. Wu are with the National Mobile Communications Research Laboratory, School of Information Science and Engineering, Southeast University, Nanjing 211189, China, and also with  Purple Mountain Laboratories, Nanjing 211111, China (e-mail: chxwang@seu.edu.cn, xiping.wu@seu.edu.cn).

S. Chen and C. Huang are with Purple Mountain Laboratories, Nanjing 211111, China; and with the National Mobile Communications Research Laboratory, School of Information Science and Engineering, Southeast University, Nanjing 211189, China (e-mail: shuaifeichen@seu.edu.cn, huangchen@pmlabs.com.cn).

E. Bj{\"o}rnson is with the Department of Computer Science, KTH Royal Institute of Technology, 164 40 Kista, Sweden (e-mail: emilbjo@kth.se).

}
}

\markboth{To appear in IEEE Communications Letters, 10.1109/LCOMM.2025.3637318}%
{Shell \MakeLowercase{\textit{et al.}}: A Sample Article Using IEEEtran.cls for IEEE Journals}

\maketitle

\begin{abstract}
This letter investigates channel estimation for ultra-massive multiple-input multiple-output (MIMO) communications. We propose a joint low-rank and sparse Bayesian estimation (LRSBE) algorithm for spatial non-stationary ultra-massive channels by exploiting the low-rankness and sparsity in the beam domain. Specifically, the channel estimation integrates sparse Bayesian learning and soft-threshold gradient descent within the expectation-maximization framework. Simulation results show that the proposed algorithm significantly outperforms the state-of-the-art alternatives under different signal-to-noise ratio conditions in terms of estimation accuracy and overall complexity.
\end{abstract}


\section{Introduction}

Employing a large number of antennas significantly improves the spectral and energy efficiencies of sixth-generation (6G) wireless communications with enhanced beamforming gain and spatial interference suppression capability \cite{6G,chen2021structured}. 
However, having such large channel dimensions poses a critical channel estimation challenge. This issue is further aggravated by the spatial non-stationary phenomenon of ultra-massive multiple-input multiple-output (MIMO) channels and becomes more prominent in the near-field region \cite{MIMO CE}. Conventional least squares and minimum mean-square-error (MSE) algorithms are ill-suited for such scenarios, as they fail to exploit channel's structural properties, i.e., low-rankness and sparsity~\cite{DOA}. Consequently, these methods cannot achieve a satisfactory trade-off between estimation accuracy and computational complexity.

Compressive sensing (CS)-based channel estimation is well-attended by leveraging the channel sparsity and recovering the channel from limited measurements through a small number of samples. 
CS algorithms can be classified into three categories: convex optimization\cite{ista}, greedy\cite{OMP}, and Bayesian algorithms\cite{zhang2013,BSBL}, emphasizing estimation accuracy, efficiency, and robustness, respectively. 
The low-rank matrix recovery-based channel estimation algorithms utilize the channel low-rankness, which means the channel matrix has a small rank. Among these, nuclear norm minimization-based approaches lead to a satisfactory estimation solution by transforming the rank minimization problem into the convex nuclear norm minimization\cite{NNM}. Tensor decomposition approaches capture structural information about the higher dimensions of the channel through tensor representations\cite{TD,TDn}.
However, estimation based on sole channel low-rankness or sparsity is suboptimal, and thus, a joint design is needed to further improve channel estimation performance.

Beam domain channel estimation is promising since the ultra-massive MIMO channel's low-rankness and sparsity are both observable in the beam domain, where signal energies are concentrated in specific beam directions due to the typically limited signal propagation paths \cite{chen2025channel}. 
In \cite{two-stage}, a two-stage channel estimation algorithm was proposed by exploiting both low-rankness and sparsity but in a separate way. Though the method can improve the estimation accuracy to some extent, it failed to address the joint optimization, causing a possible error propagation across the two stages. 
To this end, \cite{DOA} performed directions-of-arrival (DoA) tracking by Bayesian inference, assuming that stationary paths are low-rank and time-varying paths are sparse. Nevertheless, the modeling assumptions are relatively simple, and the adopted DoA tracking approach in the space domain cannot be directly applied to channel estimation due to the inhibited complexity.

To fill the research gaps, this letter proposes a high-accuracy joint low-rank and sparse Bayesian estimation (LRSBE) algorithm for ultra-massive MIMO channels. The main contributions are summarized as follows:
\begin{itemize}
    \item We propose to decompose the beam domain channel into independent low-rank and sparse components based on the visibility across the antenna array.
    \item We formulate a joint low-rank and sparse recovery problem for channel estimation optimization and solve it through Bayesian learning within the expectation-maximization (EM) framework.
    \item The proposed algorithm is compared with state-of-the-art alternatives for validation of effectiveness and efficiency.
\end{itemize}

The rest of this paper is organized as follows. Section~II introduces the beam domain channel model and analyzes its characteristics. In Section~III, the proposed LRSBE algorithm is described in detail. Section IV provides simulation results and analysis. Finally, conclusions are drawn in Section~V.

\section{System Model}

Let us consider $K$ single-antenna mobile terminals (MTs) and a base station (BS) equipped with a large uniform planar array comprising $M\!=\!{M_{\rm h}}\!\times\!{M_{\rm v}}$ half-wavelength-spaced antennas, where $M_{\rm h}$ and $M_{\rm v}$ are the numbers of antennas in the horizontal and vertical directions, respectively. 
We consider the block fading model and let ${\bf h}_k ={\rm vec}\left({\bf H}_k\right) \in  {\mathbb C}^{M}$ 
denote beam domain channel transfer function between MT $k$ and the BS in a specific coherence block, where ${\rm vec}(\cdot)$ denotes vectorization. The beam domain channel matrix ${\bf{ H}}_k\!\in\! \mathbb{C}^{{M_{\rm h}}\times{M_{\rm v}}}$ is given by
\begin{equation}\label{eq:BDCM}
    \mathbf {H}_k = \mathbf{U}_{\rm h}^{\rm H}\mathbf {\bar H}_k\mathbf{U}_{\rm v} 
\end{equation}
where ${\bf{\bar H}}_k\!\in\! \mathbb{C}^{{M_{\rm h}}\times{M_{\rm v}}}$ is the corresponding space domain channel matrix resulting from the 6G pervasive channel model (6GPCM) \cite{beam1}, by which the spatial non-stationarity of ultra-massive MIMO can be characterized by dividing the clusters into partially visible and wholly visible parts\cite{BDCM}. More details will be elaborated later in this section. Matrices $\mathbf{U}_{\rm h}\!\in\! \mathbb{C}^{{M_{\rm h}}\times{M_{\rm h}}}$ and $\mathbf{U}_{\rm v}\!\in\! \mathbb{C}^{{M_{\rm v}}\times{M_{\rm v}}}$ are horizontal and vertical discrete Fourier transform-based (DFT) beamforming matrices, respectively. 

\begin{figure}[!t]
\centering
\vspace{-1em}
\!\!\!\!\!\!\!\includegraphics[width=0.55\textwidth]{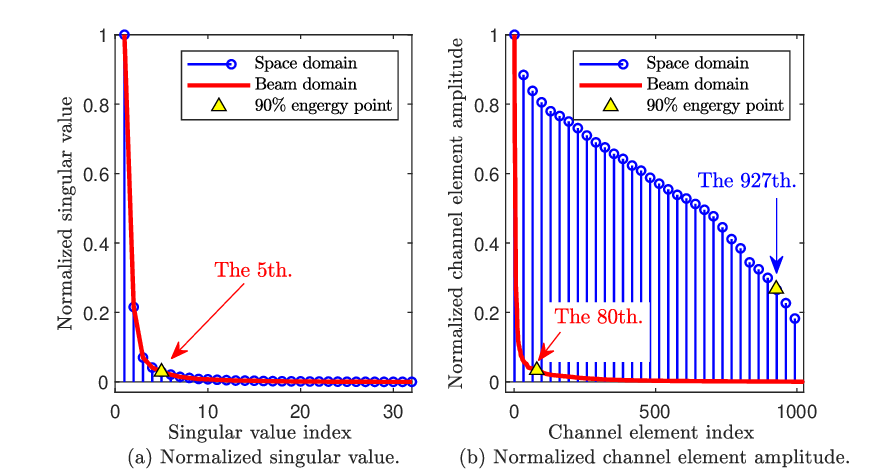}
\vspace{-1.5em}\caption{Illustration of (a) low-rankness and (b) sparsity of the space domain and beam domain channels ($M$ = 1024).}
\label{fig_1}\vspace{-1em}
\end{figure}

The space domain ultra-massive MIMO channel matrix ${\bar{\bf H}}_k$ and its beam domain representative ${\bf H}_k$ are both low-rank by having limited scattering clusters and strong spatial correlation, while only the latter is sparse due to the focusing effect of DFT in \eqref{eq:BDCM}. Fig. \ref{fig_1} quantitatively exhibits the channel low-rankness and sparsity in both the beam and space domains by showing the distributions of the singular values of the channel matrices and the amplitude of channel elements. In Fig.~\ref{fig_1}(a), most of the information is carried by a few singular values, and the low-rankness is evident in both the beam and space domains by having 5 dominant singular values constituting 90\% of the energy.
As for the channel sparsity illustrated in Fig.~\ref{fig_1}(b), channel elements are non-negligible in the space domain, and the beam domain channels consist of a few dominant elements with the remaining elements very close to zero. 
According to the above observations, we can rewrite the beam domain channel ${\bf H}_k$ as
\begin{equation}\label{eq:visibility}
\mathbf {H}_k=\mathbf {H}_k^{\rm{L}} + \mathbf {H}_k^{\rm{S}}
\end{equation}
where $\mathbf {H}_k^{\rm{L}}\!\in\! \mathbb{C}^{{M_{\rm h}}\times{M_{\rm v}}}$ and $\mathbf {H}_k^{\rm{S}}\!\in\! \mathbb{C}^{{M_{\rm h}}\times{M_{\rm v}}}$ are low-rank and sparse components, respectively.
One physical realization of \eqref{eq:visibility} is the channel decomposition according to the visibility across a huge BS antenna array [15, Eq. (29)]. In this case, $\mathbf {H}_k^{\rm{L}}$ and $\mathbf {H}_k^{\rm{S}}$ coincide with the partially visible component and wholly visible component representing the clusters of ${\bf H}_k$ seen by a subset of antennas and all antennas, respectively. This comes from the fact that the channel constructed by the wholly visible component consists of paths in the major few beam directions and thus exhibits a strong sparsity. The channel composed of the partially visible component, on the other hand, has a relatively slow amplitude change due to the presence of scatterers, and the main fluctuations are concentrated in a few columns of the channel matrix, which exhibits a strong low-rankness. 
\vspace{-0.5em}

\section{\!Joint Low-rank and Sparse Channel \!Estimation}

\subsection{Joint Optimization Problem Formulation}

Let ${\bf h} \!=\! [{\bf h}_1^{\top}\!,\ldots,\!{\bf h}_K^{\top}]^{\top}\!\in\! {\mathbb C}^{M K}$ denote the collective channel and ${\bf x}_k \!\in\! {\mathbb C}^N$ represent the pilot sequence transmitted the $k$th MT in the uplink. Here, $N\!\ll\!K$ is the number of orthogonal pilot sequences. It is worth noting that the so-called pilot contamination deteriorating the coherent transmission could occur in this case, due to the pilot reuse  among the MTs. By leveraging the sparsity of the channel, a reduced number of pilots can be designed to alleviated the impact of pilot contamination, which will be further investigated in the future work.
The received signal at BS is given by~\cite{zhang2013}
\vspace{-0.5em}\begin{equation}
\begin{aligned}
\mathbf {y} &= \mathbf{A}\mathbf{h} +\mathbf{n} 
\\
&= \mathbf{A}\left(\mathbf{h}^{\rm S}+\mathbf{h}^{\rm L}\right) +\mathbf{n} \in {\mathbb C}^{MN}
\label{transmission}
\end{aligned}
\vspace{-0.5em}\end{equation}
where ${\bf A}=[{\bf x}_1,\ldots,{\bf x}_K]\otimes {\bf I}_M \in {\mathbb C}^{MN \times MK}$ is the prior measurement matrix, ${\bf h}^{\rm S} = [({\bf h}_1^{\rm S})^{\top},\ldots,({\bf h}_K^{\rm S})^{\top}]^{\top}$ with ${\bf h}_k^{\rm S}\!=\!{\rm vec}({\bf H}_k^{\rm S})$, ${\bf h}^{\rm L} = [({\bf h}_1^{\rm L})^{\top},\ldots,({\bf h}_K^{\rm L})^{\top}]^{\top}$ with ${\bf h}_k^{\rm L}={\rm vec}({\bf H}_k^{\rm L})$, and ${\bf n} \sim \mathcal{N}_\mathbb{C}(\bf 0,\sigma^2\mathbf{I}_{MN})$ is the receiver noise with power $\sigma^2$.

The relationship between the prior distribution $P(\mathbf {h})$, the posterior distribution $P(\mathbf {h} \! \mid \! \mathbf y)$, and the likelihood function $P(\mathbf y\mid \mathbf {h})$ of channel ${\bf h}$ can be expressed as 
\vspace{-0.5em}\begin{equation}
    P(\mathbf {h}\mid \mathbf y)P(\mathbf y)= P(\mathbf y\mid \mathbf {h}) P(\mathbf {h}) \label{Bayesian}
\vspace{-0.5em}\end{equation}
according to Bayes' theorem, where the likelihood
$P(\mathbf y\mid \mathbf {h}) = P(\mathbf {y} \mid \mathbf{h}^{\rm S}+\mathbf{h}^{\rm L})$
describes the distribution of $\mathbf y$ for a given $\mathbf {h}=\mathbf{h}^{\rm S}+\mathbf{h}^{\rm L}$. 
Since the noise in the system is complex Gaussian, its distribution is
\vspace{-0.5em}\begin{equation}
P(\mathbf{n}) = \frac{1}{ \left(\pi\sigma^2\right)^{MN}} \exp \left(-\frac{\|\mathbf{n}\|_2^2}{ \sigma^2}\right)
\vspace{-0.5em}\end{equation}
where $\left\|\cdot\right\|_2$ stands for the $\ell_2$-norm of the vector. Therefore, the likelihood function also follows the complex Gaussian distribution, i.e.,
\vspace{-0.5em}\begin{equation}
\begin{aligned}
P(\mathbf y\mid \mathbf {h}) &= P(\mathbf {y} \mid \mathbf{h}^{\rm S}+\mathbf{h}^{\rm L})
\\
&=\frac{1}{\left(\pi\sigma^2\right)^{MN}} \exp \!\left(-\frac{\left\|\mathbf{y}-\mathbf{A}\left(\mathbf{h}^{\rm S}+\mathbf{h}^{\rm L}\right)\right\|_2^2}{\sigma^2}\right). \label{likelihood}
\end{aligned}
\vspace{-0.5em}\end{equation}

In order to reflect the sparsity and low-rankness while maintaining mathematical solvability, we model the prior distributions using complex Gaussian distributions with adjustable variances\cite{zhang2013, DOA}. The prior distributions of ${\bf h}^{\rm S}$ and ${\bf h}^{\rm L}$ are given by
\vspace{-0.5em}\begin{equation}
\begin{aligned}
P(\mathbf{h}^{\rm S}) = \mathcal{N}_\mathbb{C}\left(\mathbf{0}, \alpha^{-1} {\bf I}_{MK}\right)
\\
P(\mathbf{h}^{\rm L}) = \mathcal{N}_\mathbb{C}\left(\mathbf{0}, \beta^{-1} \mathbf{I}_{MK}\right) \label{pri}
\end{aligned}
\vspace{-0.5em}\end{equation}
where $\alpha$ and $\beta$ are hyperparameters controlling the levels of sparsity and low-rankness, respectively. As $\alpha$ and $\beta$ increase, the variances decrease and more elements are concentrated around zero, resulting in stronger sparsity and low-rankness. 

The posterior distribution $P({\bf h} | {\bf y})$ is derived through Bayesian learning based on the prior distribution and likelihood function\cite{zhang2013}. Assuming that channel multi-paths are independent regarding phase, delay, and amplitude without coupling or interference, $\mathbf {h}^{\rm S}$ and $\mathbf {h}^{\rm L}$ can be considered mutually independent. Substituting (\ref{likelihood}) and (\ref{pri}) into (\ref{Bayesian}), the posterior distribution of $\bf h$ can be expressed as
\vspace{-0.25em}\begin{equation}\label{eq:post}
\begin{aligned}
&P(\mathbf{h} \mid \mathbf {y}) = P(\mathbf {h}^{\rm S}, \mathbf {h}^{\rm L} \mid \mathbf {y}) \\
&\propto P(\mathbf y\mid \mathbf {h}) P(\mathbf {h}^{\rm S}) P(\mathbf {h}^{\rm L})
\\
\!\!&\propto \! \exp \! \left(\!\!-\frac{\left\|\mathbf{y}\!-\!\mathbf{A}\!\left(\mathbf{h}^{\rm S}\!+\!\mathbf{h}^{\rm L}\!\right)\right\|_2^2}{ \sigma^2} \!-\! \alpha  \left\|\mathbf{h}^{\rm S}\right\|_2^2 \!-\! \beta \left\|\mathbf{h}^{\rm L}\right\|_2^2 \right).
\end{aligned}
\vspace{-0.25em}\end{equation}
The optimal estimate of $\bf h$ can be obtained by maximizing the posterior distribution presented in \eqref{eq:post}. Since the posterior distribution is a monotonically increasing exponential function, by removing the negative sign and neglecting the constant terms, we can estimate $\bf h$ by solving
\vspace{-0.25em}\begin{equation}\label{eq:P1}
{\sf P}_1\!:\ \underset{\mathbf{h}^{\rm S}, \mathbf{h}^{\rm L}}{\min}\  \!\left\|\mathbf{y}\!-\!\mathbf{A}\left(\mathbf{h}^{\rm S}\!+\!\mathbf{h}^{\rm L}\right)\right\|_2^2 \!+ \alpha  \left\|\mathbf{h}^{\rm S}\right\|_0 \!+ \beta \ {\rm Rank} \left(\mathbf{H}^{\rm L}\right)
\vspace{-0.25em}\end{equation}
where ${\bf H}^{\rm L} = [{\bf H}_1^{\rm L},\ldots,{\bf H}_K^{\rm L}]$ is the low-rank component of the collective channel matrix with $\mathbf{h}^{\rm L}={\rm vec}\left({\bf H}^{\rm L}\right)$. Moreover, $\left\|\cdot\right\|_0$ and ${\rm Rank} (\cdot)$ represent the $\ell_0$-norm of the vector and rank of the matrix, respectively.
The first term in \eqref{eq:P1} is the data fitting term in \eqref{eq:post}, representing the estimation error. The second and third terms are the sparse penalty term and the low-rank penalty term, respectively. We minimize the estimation error while simultaneously promoting sparsity and low-rankness through the corresponding penalty terms. This unconstrained formulation is equivalent to the Lagrangian form of a constrained problem that imposes explicit bounds on the sparsity and rank. Furthermore, in order to ensure the convexity of the problem, $\ell_1$-norm is introduced to constrain the sparsity and the nuclear norm is used as a convex approximation of the rank function \cite{nuclear}.
As a result, we formulate the process of estimating the beam domain channel $\bf h$ as a joint low-rank and sparse recovery problem ${\sf P}_2$, which is given by
\vspace{-0.25em}\begin{equation}\label{eq:P2}
{\sf P}_2:\ \underset{\mathbf{h}^{\rm S}, \mathbf{h}^{\rm L}}{\min}\  \left\|\mathbf{y}-\mathbf{A}\left(\mathbf{h}^{\rm S}\!+\!\mathbf{h}^{\rm L}\right)\right\|_2^2 \!+\! \alpha  \left\|\mathbf{h}^{\rm S}\right\|_1 \!+\! \beta  \left\|{\bf H}^{\rm L}\right\|_\ast
\vspace{-0.25em}\end{equation}
where $\left\|\cdot\right\|_1$ is the $\ell_1$-norm of the vector and $\left\|\cdot\right\|_\ast$ is the nuclear norm of the matrix.

\vspace{-0.5em}
\subsection{EM Framework for Iterative Updating}

\begin{algorithm}[!t]
    \caption{The LRSBE Algorithm.}
    \label{algorithm}
    \renewcommand{\algorithmicrequire}{\textbf{Input:}}
    \renewcommand{\algorithmicensure}{\textbf{Output:}}
    \begin{algorithmic}[1]        
      \REQUIRE ${\bf y} \!\in\! {\mathbb C}^{MN}$, ${\bf A} \!\in \!{\mathbb C}^{MN \times MK}$, maximum allowed iterations $Q$.
      \ENSURE $\hat{\mathbf{h}}\in {\mathbb C}^{MK}$.  
      \STATE \textbf{Initialization:} $i=1$, $\hat{\mathbf{h}}^{\rm S}_i = \mathbf{0}$, $\hat{\mathbf{h}}^{\rm L}_i = \mathbf{0}$, $\alpha=1$, $\beta=1$;

      \WHILE{$i\le Q$}
      \STATE Update $\mathbf{r}^{\rm S} = \mathbf{y}-\mathbf{A}\hat{\mathbf{h}}^{\rm L}_i$;
      \STATE Calculate $\mu$ and $\bf \Sigma$ through \eqref{mean} and \eqref{cov};
      \STATE Calculate $\mathbf{\Gamma}$ and $\mathbf{C}$ through \eqref{corr} and \eqref{bcov};
      \STATE Update $\hat{\mathbf{h}}^{\rm S}_{i+1} = \mu$;
      \STATE Update $\mathbf{r}^{\rm L} = \mathbf{y}-\mathbf{A}\hat{\mathbf{h}}^{\rm S}_{i+1}$;
      \STATE Update $\hat{\mathbf{h}}^{\rm L}_{i+1}$ through \eqref{gradient descent} and \eqref{kernel norm constraint};
      \STATE Limit the number of singular values through \eqref{SVD};
      \STATE Combine estimation result $\hat{\mathbf{h}}_{i+1}=\hat{\mathbf{h}}^{\rm S}_{i+1}+\hat{\mathbf{h}}^{\rm L}_{i+1}$;
      \STATE Update $\alpha$ and $\beta$ through (\ref{beta});
        \IF {${\|\hat{\mathbf{h}}_{i+1}-\hat{\mathbf{h}}_{i}\|_2}/{\|\hat{\mathbf{h}}_{i}\|_2} \le 10^{-4}$}          
          \STATE break;
        \ENDIF
    \STATE $i\leftarrow i+1$;
      \ENDWHILE
      
      \RETURN $\hat{\mathbf{h}}=\hat{\mathbf{h}}^{\rm S}_{i}+\hat{\mathbf{h}}^{\rm L}_{i}$.
      
    \end{algorithmic}
\end{algorithm}
Since \eqref{eq:P2} contains non-smooth $\ell_1$-norm and the hyperparameters $\alpha$ and $\beta$ need to be dynamically updated according to the current estimation results, we employ the EM framework~\cite{zhang2013}, which is particularly suitable for probabilistic models containing hidden or unobserved variables, to solve the formulated problem $\sf P_2$.
Within the EM framework, $\mathbf{h}^{\rm S}$ and $\mathbf{h}^{\rm L}$, which is equivalent to ${\bf H}^{\rm L}$, are treated as hidden variables and $\alpha$ and $\beta$ as model parameters.
\subsubsection{E-step}
This step calculates posterior expectations of $\mathbf{h}^{\rm S}$ and $\mathbf{h}^{\rm L}$ with fixed hyperparameters. 
The nuclear norm and $\ell_1$-norm terms make \eqref{eq:P2} difficult to be directly optimized.
To this end, we propose a stepwise optimization approach that constrains the low-rankness and sparsity separately.

Considering that the channel sparse component exhibits block sparsity structure \cite{block}, $\mathbf{h}^{\rm S}$ can be processed in blocks.
Let $\mathbf{h}^{\rm S} \!\!=\! [ (\mathbf{h}^{\rm S}_{1})^{\top}, \!\ldots,\! (\mathbf{h}^{\rm S}_{G})^{\top}]^{\top}$,
where $G$ is the number of blocks, $L$ is the length of the single block, and $\mathbf{h}^{\rm S}_{k}$ is the $k$th block. The block length is designed based on the antenna dimension. 
Fixing $\mathbf{h}^{\rm L}$, the optimal $\mathbf{h}^{\rm S}$ is achieved by solving
\vspace{-0.5em}\begin{equation}
{\sf P}_3: \ \underset{\{\mathbf{h}^{\rm S}_{k}\}}{\min}\  \left\|\mathbf{r}^{\rm S}-\mathbf{A}_{k}\mathbf{h}^{\rm S}_{k}\right\|_2^2 + \alpha \sum\nolimits_{k=1}^G \left\|\mathbf{h}^{\rm S}_{k}\right\|_1
\vspace{-0.5em}\end{equation}
where $\mathbf{r}^{\rm S} = \mathbf{y}-\mathbf{A}\mathbf{h}^{\rm L}$ denotes the sparse residual of the channel and $\mathbf{A}_{k}\in {\mathbb C}^{MN \times L}$ is an extracted measurement matrix in which the columns are associated with the $k$th block. Utilizing the block sparse Bayesian learning \cite{BSBL}, the posterior mean and covariance of $\mathbf{h}^{\rm S}$ are
\vspace{-0.5em}\begin{align}
\label{mean} \mu &= \mathbf{\Gamma} \mathbf{A}^{\rm H} (\mathbf{A}\mathbf{\Gamma}\mathbf{A}^{\rm H} + \sigma^{2} \mathbf{I})^{-1} \mathbf{r}^{\rm S}\\
 \label{cov} \mathbf{\Sigma} &= \mathbf{\Gamma} - \mathbf{\Gamma} \mathbf{A}^{\rm H} (\mathbf{A}\mathbf{\Gamma}\mathbf{A}^{\rm H} + \sigma^{2} \mathbf{I})^{-1} \mathbf{A} \mathbf{\Gamma} 
\vspace{-0.85em}\end{align}
where $\mathbf{\Gamma}\!=\!{\rm diag} (\gamma_1\mathbf{C}_1, \dots, \gamma_G\mathbf{C}_G)$ represents the block covariance matrix, with $\mathbf{C}_k$ and $\gamma_k$ being the correlation matrix and weight of $\mathbf{h}^{\rm S}_{k}$, respectively. 
Since assigning each block with an exclusive $\mathbf{C}_k$ easily leads to overfitting \cite{zhang2013}, the correlation matrices are set to be identical. We have
\vspace{-0.3em}\begin{align}
    \mathbf{C}_k &= \mathbf{C}= \frac{1}{G} \sum\nolimits_{k=1}^G \frac{\mathbf{\Sigma}_k + \mu_k \mu_k^{\rm H}}{\gamma_k},\quad \forall k \label{corr}\\
    \gamma_k&= \frac{1}{L} \operatorname{tr}(\mathbf{C}^{-1}(\mathbf{\Sigma}_k + \mu_k \mu_k^{\rm H})) \cdot \frac{1}{1+\alpha}. \label{bcov}
\vspace{-1.em}\end{align}
Since the block sparsity is susceptible to noise, $L$ is set to be a variable to achieve adaptive adjustment for different scenarios.

The low-rank component ${\bf h}^{\rm L}$ can be obtained by solving

\vspace{-0.8em}\begin{equation}\label{eq:P4}
{\sf P}_4: \     \underset{\mathbf{h}^{\rm L}}{\min}\ \left\|\mathbf{r}^{\rm L}-\mathbf{A}\mathbf{h}^{\rm L}\right\|_2^2 + \beta\left\|{\bf H}^{\rm L}\right\|_\ast
\vspace{-0.5em}\end{equation}
where $\mathbf{r}^{\rm L} \!= \!\mathbf{y}\!-\!\mathbf{A}\mathbf{h}^{\rm S}$ represents the low-rank residual of the channel. 
To minimize the first term of \eqref{eq:P4}, its gradient is calculated as
\vspace{-0.25em}\begin{equation}
    \nabla_{\mathbf{h}^{\rm L}}\left(\left\|\mathbf{r}^{\rm L}-\mathbf{A}\mathbf{h}^{\rm L}\right\|_2^2\right)=-\mathbf{A}^{\rm H}\left(\mathbf{r}^{\rm L}-\mathbf{A}\mathbf{h}^{\rm L}\right). \label{gra}
\vspace{-0.25em}\end{equation}
Derived by the gradient descent formula, the initial estimates for the $i$th iteration is
\begin{equation}
    {\mathbf{h^{\prime}}^{\rm L}_{i+1}} = \mathbf{h}^{\rm L}_{i}+\frac{1}{T} \mathbf{A}^{\rm H}\left(\mathbf{r}^{\rm L}-\mathbf{A} \mathbf{h}^{\rm L}_{i}\right) \label{gradient descent}
\end{equation}
where $T=\lambda_{\max }\left(\mathbf{A}^{\rm H} \mathbf{A}\right)$ is the step length, set to be the maximum eigenvalue of $\mathbf{A}^{\rm H} \mathbf{A}$ to ensure convergence. 

To minimize the regularization term $\beta\left\|{\bf H}^{\rm L}\right\|_\ast$ at the same time, we utilize a soft-thresholding operation~\cite{ista} to impose a regularization constraint, i.e.,
\begin{equation}
    h^{\rm L}_{k,i+1}=\operatorname{sign}({h^{\prime}}^{\rm L}_{k,i+1}) \cdot \max (0,|{h^{\prime}}^{\rm L}_{k,i+1}|-\tau) \label{kernel norm constraint}
\end{equation}
where $\operatorname{sign}\left(\cdot\right)$ is the sign function, $h^{\rm L}_{k,i+1}$ and ${h^{\prime}}^{\rm L}_{k,i+1}$ represent the $k$th elements of $\mathbf{h}^{\rm L}$ and $\mathbf{h^{\prime}}^{\rm L}$ obtained in the $i$th iteration, respectively. The threshold $\tau \!=\! \frac{\beta}{2 T}$ controls the level of low-rankness. As the low-rank parameter $\beta$ increases, more entries of ${\bf h}^{\rm L}$ are ``compressed" to zero. 

After performing singular value decomposition on ${\bf H}^{\rm L}$, we apply soft thresholding to the singular values as
\begin{equation}
    \sigma_k = \max \left(\sigma_k-\frac{\sqrt{\beta}}{2}, 0\right). \label{SVD}
\end{equation}
As a result, the small singular values are removed.

\begin{figure}[!t]
\vspace{-1.5em}\centering
\!\!\!\!\!\!\!\includegraphics[width=0.55\textwidth]{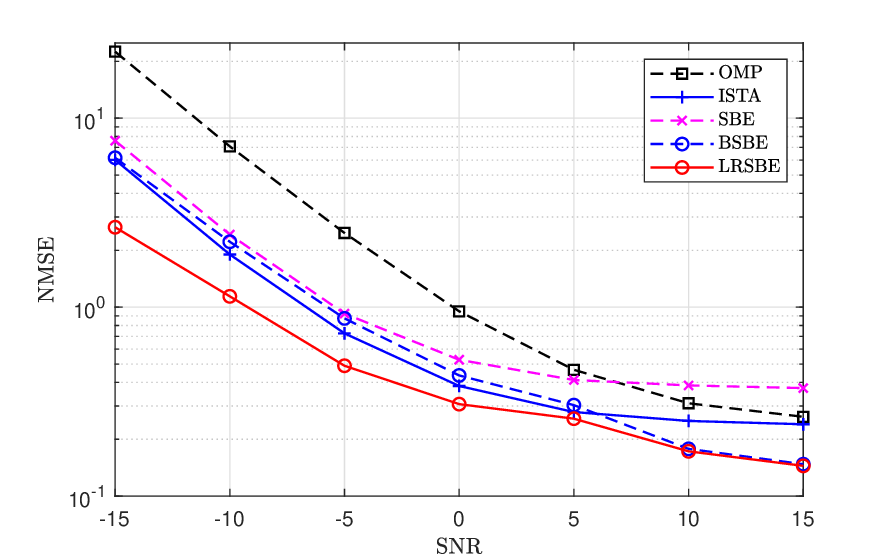}
\vspace{-2em}\caption{Comparison of average NMSEs versus SNR for different algorithms ($M=256, N=5$).}
\label{fig_2}\vspace{-1em}
\end{figure}

\subsubsection{M-step}
Based on the current estimations of $\mathbf{h}^{\rm S}$ and $\mathbf{h}^{\rm L}$, the M-step adjusts the hyperparameters $\alpha$ and $\beta$  by maximizing the likelihood functions, thus dynamically adapting to different channel conditions. The log-likelihood functions of $\mathbf{h}^{\rm S}$ and $\mathbf{h}^{\rm L}$ with respect to $\alpha$ and $\beta$ are
\vspace{-0.5em}\begin{equation}
\begin{aligned}
	\log P\left(\mathbf{h}^{\rm S}\right)&=-\alpha\left\|\mathbf{h}^{\rm S}\right\|_2^2+MK \log \alpha\\
	\log P\left(\mathbf{h}^{\rm L}\right)&=-\beta\left\|\mathbf{h}^{\rm L}\right\|_2^2+MK\log \beta.
    \label{pri_beta}
\end{aligned}
\end{equation}
Considering the posterior expectations of $\mathbf{h}^{\rm S}$ and $\mathbf h^{\rm L}$, we have
\vspace{-0.5em}\begin{equation}
    \begin{aligned}
        \alpha&=\frac{MK}{\left\|\mathbf{y}-\mathbf{A}\left(\mathbf{h}^{\rm S} + \mathbf{h}^{\rm L}\right)\right\|_{\rm 2}^2+\frac{1}{\alpha} \sum_k\left(1-\theta_k\right)} \\
      \beta&=\frac{MK}{\left\|\mathbf{h}^{\rm L}\right\|_{\rm 2}^2 + \operatorname{tr}(\boldsymbol{\Sigma}^{\rm L})}\label{beta}
    \end{aligned}
\end{equation}
where $\theta_k={(\mu^{\rm S}_{k})}^2 / \Sigma^{\rm S}_{k k}$ represents the posterior weight for the $k$th entry of $\mathbf{h}^{\rm S}$, in which $\mu^{\rm S}_{k}$ and $\Sigma^{\rm S}_{k k}$ are posterior mean and covariance of the $k$th component of $\mathbf{h}^{\rm S}$, respectively.
Besides, $\boldsymbol{\Sigma}^{\rm L}$ denotes the posterior covariance of $\mathbf{h}^{\rm L}$.

This algorithm handles low-rankness and sparsity with nuclear norm and $\ell_1$-norm regularization, respectively. The procedure is summarized in Algorithm \ref{algorithm}.
\vspace{-0.5em}
\section{Simulation Results}
An ultra-massive MIMO system consisting of $K=10$ MTs is considered in an urban microcell scenario where the BS is equipped with $M$ antennas, $M\in\{16\times 16 = 256, 32\times 32 = 1024\}$. The system operates at a carrier frequency of 11 GHz and the channel is generated according to \cite{BDCM}. 
The simulations are performed in MATLAB R2023b on a computer with Intel Core i7-12700 CPU @ 2.1 GHz and 16 GB of RAM.
We conduct a comparative analysis to evaluate the performance of our proposed LRSBE algorithm. The analysis involves four representative comparative algorithms, which cover three categories of the CS algorithm. Specifically, the comparative algorithms selected include the greedy algorithm orthogonal matching pursuit (OMP) \cite{OMP}, the convex optimization-based iterative shrinkage-threshold algorithm (ISTA) \cite{ista}, the sparse Bayesian learning estimation (SBE) algorithm \cite{zhang2013}, and our previously proposed block SBE (BSBE) algorithm\cite{BSBL}. The normalized MSE (NMSE) $\frac{1}{K} \sum_{k=1}^K \mathbb E \{ {{\|\mathbf{h}_k - \hat{\mathbf{h}}_k \|_2^2} / {\|\mathbf{h}_k\|_2^2}} \}$ is employed to evaluate the estimation accuracy.

Fig. \ref{fig_2} demonstrates that the proposed LRSBE outperforms other algorithms regarding NMSE under different signal-to-noise ratio (SNR) conditions. Since greedy algorithms are vulnerable to noise, OMP performs relatively unsatisfactorily under low SNR. In contrast, SBE has lower NMSE due to the self-inflicted prior assumption, with BSBE improving the estimation accuracy by further utilizing the block structure of channels. By using $\ell_1$-norm regularization, ISTA can suppress the effect of noise so that its accuracy is even better than BSBE but encounters a bottleneck due to a fixed soft threshold. However, these algorithms only utilize the channel sparsity while ignoring the channel low-rankness in the low SNR region that can be exploited to overcome the influence of noise. By integrating the low-rankness and sparsity simultaneously, the proposed LRSBE further exploits the block sparsity of the channel after 5 dB, thus significantly improves the estimation accuracy with a 3~dB (\! $\approx$\! 50\%) NMSE reduction compared to the best benchmark ISTA when SNR = -15 dB and a 1~dB (\! $\approx$\! 20\%) reduction compared to the best benchmark BSBE when SNR = 15 dB.

\begin{figure}[!t]
\centering
\!\!\!\!\!\!\!\includegraphics[width=0.55\textwidth]{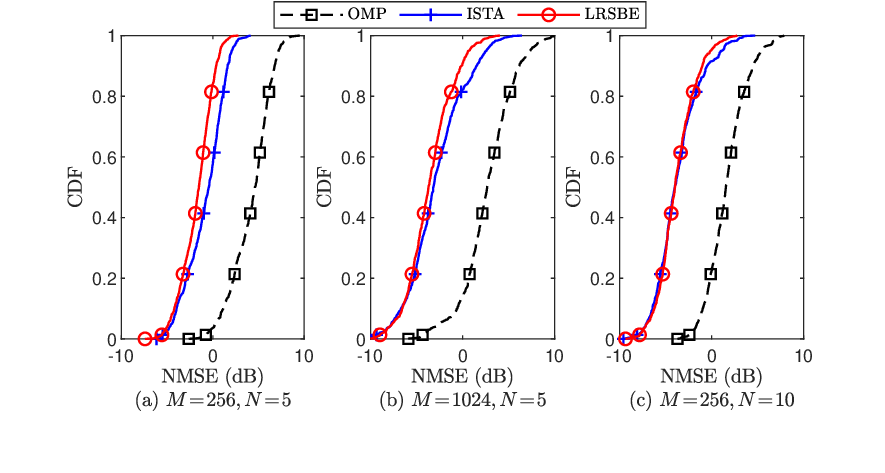}
\vspace{-3em}\caption{CDFs of NMSEs with (a) $M= 256, N = 5$, (b) $M= 1024, N = 5$ and (c) $M = 256, N = 10$ (SNR = -10 dB).} 
\label{fig_3}\vspace{-1em}
\end{figure}

\begin{figure}[!t]
\centering
\!\!\!\!\!\!\!\includegraphics[width=0.55\textwidth]{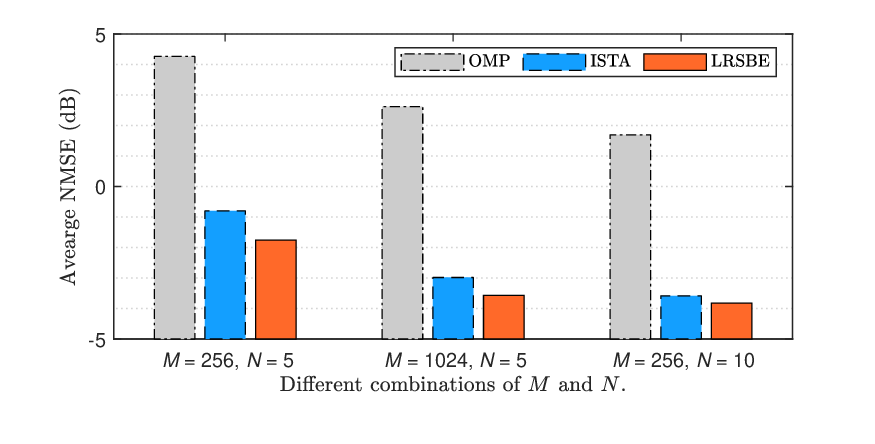}
\vspace{-3.5em}
    \caption{Average NMSEs with different combinations of $M$ and $N$ (SNR = -10 dB).} 
\label{fig_4}\vspace{-1.em}
\end{figure}

Moreover, Fig. \ref{fig_3} and Fig. \ref{fig_4} evaluate the NMSE of the proposed LRSBE with the greedy algorithm OMP and convex-based algorithm ISTA, varying the number of pilots and antenna dimensions. 
Fig. \ref{fig_3}(a), (b), and (c) display the cumulative distribution functions (CDFs) of NMSEs with three combinations of numbers of pilots and antenna dimensions, 
revealing that the estimation accuracy is improved by equipping with more antennas and utilizing more pilots.
Quantitatively, Fig.~\ref{fig_4} demonstrates the average NMSEs of the aforementioned three cases. Compared to the LRSBE with $M\!=\!256$ and $N\!=\!5$, the LRSBE with $M\!=\!1024$ and $N\!=\!5$ enjoys an almost 2 dB reduction of average NMSE through increasing the size of the antenna array. Moreover, increasing the number of pilots from 5 to 10 leads to a further 2 dB reduction in average NMSE, effectively mitigating pilot contamination when $M\!=\!256$.

In terms of the complexity per iteration of the proposed LRSBE, it is concentrated on the posterior mean and covariance operations, where the inversion makes the single iteration complexity ${\mathcal O}(M^3)$, which is the same as the SBE and BSBE~\cite{zhang2013, BSBL}. The complexity of the OMP and ISTA is ${\mathcal O}(MN)$ \cite{ista, OMP}. Fortunately, the fast convergence of the proposed LRSBE ensures low overall complexity with only a few iterations, whereas the high estimation accuracy of the BSBE is achieved at the expense of high overall complexity. This is clearly reflected in Table \ref{complexity}, which presents the average runtime and number of iterations required to meet the specified performance criteria. Here, the value of SNR is set to be 10 dB, which can ensure a stable connectivity of wireless communications. Meanwhile, to comprehensively measure the complexity of all algorithms, we adopt a precision requirement of $5\cdot10^{-1}$. As shown in Table~\ref{complexity}, the average iteration times of the proposed LRSBE is reduced by more than 80\% compared to the BSBE, which leads to a 90\% reduction of complexity. Despite the slight decrease in complexity of ISTA compared to the LRSBE, the estimation accuracy is much lower than that of LRSBE.
\vspace{-0.5em}
\section{Conclusions}

In this letter, we have proposed a LRSBE algorithm for channel estimation in ultra-massive MIMO systems. 
We have formulated a joint low-rank and sparse recovery problem and solved it through Bayesian learning within the EM framework, where the beam domain channel is decomposed into independent low-rank and sparse components.
Simulation results have exhibited significant accuracy advantages of the proposed LRSBE algorithm compared to state-of-the-art alternatives, thanks to the exploitations of channel low-rankness and sparsity at low and high SNR regions, respectively. In addition, the complexity of the LRSBE is dramatically reduced compared to other high-precision channel estimation algorithms.

\renewcommand\arraystretch{1.5}
\begin{table}[t!]
  \centering
  \fontsize{9}{12}\selectfont
  \caption{Average Runtime of The Considered Algorithms to Meet\\ NMSE = $5\cdot10^{-1}$ ($M=256$, $N=5$, and SNR = 10 $\rm dB$). }
  \label{tab:paremeter}
\vspace{-0.5em}    \begin{tabular}{|p{2.3cm}<{\centering}| p{0.8cm}<{\centering}| p{0.7cm}<{\centering}|p{0.7cm}<{\centering} | p{0.8cm}<{\centering}|p{0.8cm}<{\centering} |}
    \hline
    \bf Scheme  & OMP  & ISTA  & SBE & BSBE  & \!LRSBE \cr\hline
    \bf \!Iteration \!numbers\!\!& 601 &43 & 14 & 36 & 5\cr\hline
    \bf Runtime (s)\! & \!\!108.482 & \!13.823 & \!54.392 & \!\!229.173 & 23.965\cr\hline
    \end{tabular} \label{complexity}
\vspace{-1.5em}\end{table}

\vfill

\end{document}